\documentstyle[emulateapj]{article}

\submitted{to be published in the Astrophysical Journal, Letters}

\lefthead{Hachisu and Kato}
\righthead{Second Peak of T Coronae Borealis Outbursts}

\begin{document}

\title{A NEW INTERPRETATION FOR THE SECOND PEAK OF T CORONAE BOREALIS 
OUTBURSTS:  A TILTING DISK AROUND A VERY MASSIVE WHITE DWARF}

\author{Izumi Hachisu}
\affil{Department of Earth Science and Astronomy, 
College of Arts and Sciences, University of Tokyo,
Komaba, Meguro-ku, Tokyo 153-8902, Japan; hachisu@chianti.c.u-tokyo.ac.jp}
\and
\author{Mariko Kato}
\affil{Department of Astronomy, Keio University, 
Hiyoshi, Kouhoku-ku, Yokohama 223-8521, Japan; mariko@educ.cc.keio.ac.jp}




\begin{abstract}
A new interpretation for the second peak of T Coronae Borealis 
(T CrB) outbursts is proposed based on a thermonuclear runaway (TNR) 
model.  The system consists of a very massive white dwarf (WD) 
with a tilting accretion disk and a lobe-filling red-giant. 
The first peak of the visual light curve of T CrB outbursts is well 
reproduced by the TNR model on a WD close to the Chandrasekhar 
mass ($M_{\rm WD} \gtrsim 1.35 ~M_\odot$), while  
the second peak is reproduced by the combination of 
the irradiated M-giant and the irradiated tilting disk.
The derived fitting parameters are the WD mass
$M_{\rm WD} \sim 1.35 ~M_\odot$, the M-giant companion mass
$M_{\rm RG} \sim 0.7 M_\odot$ ($0.6-1.0 M_\odot$ is acceptable),
the inclination angle of the orbit $i \sim 70 \arcdeg$, and 
the tilting angle of the disk $i_{\rm prec} \sim 35 \arcdeg$.
These parameters are consistent with the recently derived 
binary parameters of T CrB.
\end{abstract}




%

\section{INTRODUCTION}
      T Coronae Borealis (T CrB) is one of the well observed recurrent 
novae and characterized by a secondary, fainter maximum occurring 
$\sim 100$ days after the primary peak.  
Historically, T CrB bursted twice, in 1866 and 1946, 
with the light curves very similar each other (e.g., \cite{pet46d}).  
Large ellipsoidal variations in the optical light curves
during quiescent phase suggest that an M3$-$4 red-giant component 
fills its Roche lobe (e.g., \cite{lei97}; \cite{sha97}; \cite{bel98}).
There have been debates on the nature of the hot component 
of this binary system. Webbink (1976) and Webbink et al. (1987) proposed 
an outburst mechanism of T CrB based on their main-sequence accretor model.
This accretion event model was investigated 
further by 3D numerical simulations (\cite{can92}; \cite{ruf93}).
\par
     However, Selvelli, Cassatella, \& Gilmozzi (1992) 
have been opposed to the main-sequence accretor model from their
analysis of {\it IUE} data in its quiescent state, which 
indicates the existence of a mass-accreting white dwarf (WD).
The detection of X-rays and the presence of flickering in
the optical light curves are also naturally explained in terms 
of accretion onto a WD.
Their estimated mass accretion rate in quiescent state is very high 
($\dot M_{\rm acc} \sim 2.5 \times 10^{-8} M_\odot$ yr$^{-1}$) 
and is exactly required by the thermonuclear runaway (TNR) theory 
to produce a TNR event every 80 yr on a massive ($\gtrsim 1.3 M_\odot$) 
WD.  Thus, the 1866 and 1946 outbursts can be interpreted 
in terms of a TNR event on a very massive WD. 
\par
     Rapid decline rates of the light curves indicate 
a very massive WD close to the Chandrasekhar limit, 
$M_{\rm WD} \sim  1.37-1.38 M_\odot$ (\cite{kat95}, 1999).
Assuming solar composition of the WD envelope,
Kato calculated nova light curves 
for WD masses of 1.2, 1.3, 1.35 and 1.377 $M_\odot$ and  
found that the light curve of the $1.377 M_\odot$ model 
is in better agreement with the observational light curve 
of T CrB than the other lower mass models.  
\par
     Recently, other observational supports for a massive WD
in T CrB have been reported.
Belczy\'nski and Mikolajewska (1998) derived a
permitted range of binary parameters, 
$M_{\rm WD}= 1.2 \pm 0.2 M_\odot$, 
from amplitude of the ellipsoidal
variability and constraints from the orbital solution of M-giants. 
In Shahbaz et al. (1997), a massive WD of 
$M_{\rm WD}= 1.3-2.5 M_\odot$ is suggested
from the infrared light curve fitting.  Combining these two permitted 
ranges of the WD mass in T CrB, we may conclude that 
a mass of the WD is between $M_{\rm WD}= 1.3-1.4 M_\odot$, 
which is very consistent with the light curve analysis 
$M_{\rm WD} \sim 1.37-1.38 M_\odot$ by Kato (1999).
\par
     The secondary maximum in outbursts is not generally 
observed in classical novae or in other recurrent novae.  
Selvelli et al. (1992) suggested a possibility of irradiation 
by a stationary shell around the system, although the presence
of the shell is just a speculation.  In this Letter, 
we propose another possibility of the second peak: 
irradiation by a tilting accretion disk around a massive WD. 
\par
     The main results of our analysis are:
1) the first peak is naturally reproduced by the fast developing
photosphere of the WD envelope based on the TNR model incorporated
with a very massive WD ($M_{\rm WD} \sim 1.35 M_\odot$).
2) the second peak is not fully reproduced by an irradiated M-giant 
model as simply estimated by Webbink et al. (1987);
3) the second peak can be well reproduced if we introduce 
an irradiated tilting accretion disk around the WD in addition to 
the irradiated M-giant companion.  Such tilting instabilities of 
an accretion disk have been suggested by 
Pringle (1996) for central stars as luminous as 
the Eddington limit or more (radiation-induced instability).  
Because the maximum luminosity of T CrB outbursts exceeds the Eddington 
limit (e.g., \cite{sel92}), the radiation-induced instability 
may work well in T CrB outbursts.

\section{THEORETICAL LIGHT CURVES}
     Our model is graphically shown in Figure \ref{vmaxfig}.  
The visual light is contributed to by three components of the system: 
1) the WD photosphere, 
2) the M-giant photosphere, and
3) the accretion disk surface.
\placefigure{vmaxfig}

\subsection{Decay Phase of Novae}
     In the TNR model, WD envelopes 
expand greatly as large as $\sim 100 ~R_\odot$ or more and 
then the photospheric radius gradually shrinks to the original 
size of the white dwarfs (e.g., $\sim 0.004 ~R_\odot$ for 
$M_{\rm WD}= 1.35 ~M_\odot$). 
The optical luminosity reaches its maximum 
at the maximum expansion of the photosphere and then 
gradually darkens to the level in quiescent phase.
Since the WD envelopes reach a steady-state 
in the decay phase of novae (e.g., \cite{kat94}),
we are able to treat the development of the envelope 
by a unique sequence of steady-state solutions with 
different envelope masses ($\Delta M$) 
as shown by Kato \& Hachisu (1994). 
\par
     We have calculated such sequences 
for WDs with various masses of $M_{\rm WD}= 0.6$, 0.7, 
0.8, 0.9, 1.0, 1.1, 1.2, 1.3, 1.35 and $1.377 M_\odot$
and obtained the optical light curves for the decay phase 
of TNR events.
Here, we choose $1.377 M_\odot$ as a limiting mass just below
the mass at the SN Ia explosion in W7 ($1.378 M_\odot$, 
\cite{nom84}).
We have used the updated OPAL opacity (\cite{igl96}), which  
has a strong peak near $\log T \sim 5.2$ about $20-30$\% 
larger than that of the original OPAL opacity (\cite{rog92}) 
which was used in Kato \& Hachisu (1994).  
The numerical method and various assumptions are 
the same as those in Kato \& Hachisu (1994).
It should be noted here that optically thick winds blow when
the WD envelope expands and the photospheric temperature 
decreases below $\log T_{\rm ph} \sim 5.5$.  
\par
     Each wind solution is a unique function of the envelope mass
$\Delta M$ if the WD mass is given.  The envelope mass 
is decreasing due to the wind mass loss 
at a rate of $\dot M_{\rm wind}(\Delta M)$ 
and hydrogen shell burning 
at a rate of $\dot M_{\rm nuc}(\Delta M)$, i.e., 
\begin{equation}
{{d} \over {d t}} \Delta M = \dot M_{\rm acc} - 
\dot M_{\rm wind} - \dot M_{\rm nuc},
\label{dmdt_envelope_mass}
\end{equation}
where $\dot M_{\rm acc}$ is the mass accretion rate to the WD.
Integrating equation (\ref{dmdt_envelope_mass}), 
we follow the development of the envelope mass $\Delta M$ 
and obtain physical quantities such as 
the photospheric temperature $T_{\rm ph}$, photospheric radius
$R_{\rm ph}$, photospheric velocity $v_{\rm ph}$, wind mass loss 
rate $\dot M_{\rm wind}$, and nuclear burning rate
$\dot M_{\rm nuc}$.
When the envelope mass decreases to below the critical mass,
the wind stops and after that the envelope mass is decreased 
only by nuclear burning.

\subsection{White Dwarf Photosphere}
     We have assumed a black-body photosphere of the white 
dwarf envelope.  After the optical peak, the photosphere shrinks 
with the envelope mass being blown off in the wind, 
and the photospheric temperature increases with the visual light 
decrease because the main emitting region moves blueward.  
Based on our wind solutions, we have obtained visual magnitude 
of the WD photosphere with a window function given by 
Allen (1973).   The photospheric surface is divided into 
16 pieces in the latitudinal angle ($\Delta \theta = \pi/16$) 
and into 32 pieces in the longitudinal angle 
($\Delta \phi = 2 \pi/32$) as shown in Figure \ref{vmaxfig}. 
Then, the contribution of each piece 
is summed up by considering the inclination angle to the viewer.
A linear limb-darkening law (the coefficient is $x=0.95$,
see \cite{bel98}; \cite{wil90}) is incorporated into the calculation.

\subsection{Companion M-giant Photosphere}
     To construct a light curve, we have also included 
the contribution of the companion star
irradiated by the WD photosphere.  
The surface of the companion star is assumed to fill the inner 
critical Roche lobe as shown in Figure \ref{vmaxfig}.
Numerically dividing the latitudinal angle into 
32 pieces ($\Delta \theta = \pi/32$) and the longitudinal angle into
64 pieces ($\Delta \phi = 2 \pi/64$), we have also summed up
the contribution of each area considering the inclination angle 
to the viewer by assuming the same limb-darkening law 
as the WD photosphere, but 
we neglect the gravity-darkening effect of the companion star 
because the hemisphere to the WD is heated up 
by the irradiation.
Here, we assume that 50\% of the absorbed energy 
is reemitted from the hemisphere of the companion with a black-body 
spectrum at a local temperature.
The original (non-irradiated) photospheric temperature 
of the companion star is assumed to be $T_{\rm ph, RG} = 3500$ K 
(\cite{bel98}).  If the accretion disk around the WD
blocks the light from the WD photosphere, it makes 
a shadow on the surface of the companion star.  Such an effect is also 
included in our calculation.  
\par
     The orbit of the companion star is assumed to be circular. 
The light curves are calculated for five cases of the companion mass, 
i.e., $M_{\rm RG}= 0.6$, 0.7, 0.8, 0.9, and $1.0 M_\odot$, 
as derived by Belczy\'nski \& Mikolajewska (1998).
Since we obtain similar light curves for all of these five masses,
we show here only the results for $M_{\rm RG}= 0.7 M_\odot$.
In this case, the separation is $a= 199.3 R_\odot$, the effective radius 
of the inner critical Roche lobe for the WD component is 
$R_1^*= 87.0 R_\odot$, the effective 
radius of the M-giant companion star is $R_2= R_2^*= 64.5 R_\odot$.

\subsection{Accretion Disk Surface}
     We have included the luminosity coming from the accretion disk 
irradiated by the WD photosphere when the accretion disk
reappears a few to several days after the optical maximum.
The surface of the accretion disk 
absorbs photons and reemits in the same way as the companion does.
Here, we assume that the accretion disk surface emits photons as 
a black-body at a local temperature of the heated surface. 
We assume further that 25\% of the absorbed energy is 
emitted from the surface, while 
the other is carried into interior of the accretion disk and 
eventually brought into the WD.  The temperature of
the outer edge is assumed to be $T_{\rm disk}= 2000$ K, 
which is never irradiated by the WD photosphere. 
\par
     An axisymmetric accretion disk with a thickness given by 
\begin{equation}
h = \beta R_{\rm disk} \left({{\varpi} 
\over {R_{\rm disk}}} \right)^2,
\label{flaring-up-disk}
\end{equation}
is assumed, 
where $h$ is the height of the surface from the equatorial plane,
$\varpi$ the distance on the equatorial plane 
from the center of the WD, 
$R_{\rm disk}$ the outer edge of the accretion disk,
and $\beta$ is a numerical factor showing the degree of thickness.
Here, we assume $\beta= 0.01$ during the strong wind phase because 
the flaring-up edge of the accretion disk is blown in the wind, but
it increases to $\beta= 0.15$ after the wind stops (\cite{sch97}).
The surface of the accretion disk is divided into 16 pieces
logarithmically in the radial direction and into 32 pieces evenly in
the azimuthal angle as shown in Figure \ref{vmaxfig}.
The outer edge of the accretion disk is also divided into 32 pieces
by rectangles. 
\par
     The luminosity of the accretion disk
depends strongly on both the thickness of $\beta$ and the radius of 
$R_{\rm disk}$.  
The size of the accretion disk is also assumed to be given by
\begin{equation}
R_{\rm disk} = \alpha R_1^*,
\label{accretion-disk-size}
\end{equation}
where $R_1^*$ is the effective radius of the inner critical
Roche lobe given by Eggleton's (1983) formula.
The viscous heating is neglected because it is rather smaller 
than that of the irradiation effects.
\par
     We assume that the disk is tilting due to Pringle's (1996) 
mechanism.  
The tilting disk is introduced by inclining the above disk 
by degree of $i_{\rm prec}= 35 \arcdeg$ with
a precessing angular velocity of 
\begin{equation}
\Omega_{\rm prec}= \gamma \Omega_{\rm orb}.
\end{equation}
The initial phase of precession is assumed 
to be $\phi_0= -170 \arcdeg$ at the
epoch of spectroscopic conjunction with the M-giant in front,
i.e., JD 243 1931.05 $+$ 227.67$E$ at $E=0$ (\cite{lin88}).
\placefigure{vmag1350va1}

\section{RESULTS}
     To fit the first peak of the outburst light curve,
we have calculated four cases of $V$-magnitude light curves
with four different WD masses, i.e., 
$M_{\rm WD}= 1.377 ~M_\odot$, $1.35 ~M_\odot$, 
$1.3 ~M_\odot$, and $1.2 ~M_\odot$ and found that 
both the $1.377 M_\odot$ and $1.35 M_\odot$ light curves 
are in much better agreement with the observational one 
than the other less massive ones as already shown by
Kato (1995, 1999). 
Therefore, we adopt the $M_{\rm WD}= 1.35 ~M_\odot$ in this Letter.
\par
     The optically thick wind stops about 100 days after 
the luminosity peak, around 1960 (JD 2430000+) 
as shown in Figure \ref{vmag1350va1}.  Here, we assume that
the mass accretion rate ($\dot M_{\rm acc}$) to the WD 
is increased to a few 
times $10^{-7} M_\odot$ yr$^{-1}$, because the disk is heated up
by the WD photosphere (also once engulfed by the WD photosphere).
The hydrogen shell-burning vanishes 120 days after the peak 
when $\dot M_{\rm acc} < 1 \times 10^{-7} M_\odot$ yr$^{-1}$,
while it continues for rather long period
when $\dot M_{\rm acc} > 4 \times 10^{-7} M_\odot$ yr$^{-1}$ as 
shown in Table \ref{tbl-1}.
\par
     The upper panel in Figure \ref{vmag1350va1} shows 
the theoretical $V$-magnitude light curve (solid line)
of the WD photosphere with 
$\dot M_{\rm acc}= 4 \times 10^{-7} M_\odot$ yr$^{-1}$ 
and the companion without irradiation together with the observational 
points (open circles; taken from \cite{pet46a}, b, c, d).  
Two arrows indicate
epochs of the spectroscopic conjunction with the M-giant in front
(\cite{lin88}).  The ellipsoidal light variation is clearly shown 
in the later phase.   
The second panel depicts the $V$-magnitude including
the effects of the companion irradiated by the WD.  
As already suggested by Webbink et al. (1987), this cannot 
well reproduce the second peak of the light curves.
The third panel shows the theoretical light curve 
further including the tilting accretion disk irradiated by
the WD photosphere.  It fully reproduces the observational
light curve if we choose the fitting parameters of
$\alpha= 0.7$, $\beta= 0.15$, $\gamma= 9/8$, $i_{\rm prec}= 35\arcdeg$,
and $\phi_0= -170 \arcdeg$.
\placetable{tbl-1}

\section{DISCUSSION}
     The radiation induced instability of the accretion disk sets in
if the condition
\begin{eqnarray}
{{\dot M_{\rm acc}} \over {1 \times 10^{-7}~M_\odot \mbox{~yr}^{-1}}}
& \lesssim & 2
\left( {{R_{\rm disk}} \over {50 ~R_\odot}} \right)^{1/2}
\left( {{L_{\rm bol}} \over {2 \times 10^{38} 
\mbox{~erg~s}^{-1}}} \right) \cr
&\times & \left( {{R_{\rm WD}} \over {0.004 ~R_\odot}} \right)^{1/2}
\left( {{M_{\rm WD}} \over {1.35 ~M_\odot}}\right)^{-1/2}
\label{radiation_condition}
\end{eqnarray}
is satisfied (\cite{sou97}).
Selvelli et al. (1992) estimated the accretion rate of T CrB as
$\dot M_{\rm acc} \sim 0.25 \times 10^{-7} M_\odot$ yr$^{-1}$, 
which meets condition
(\ref{radiation_condition}).  Therefore, we can expect the radiation
induced instability in T CrB system.  The growth timescale of warping
is estimated by Pringle (1996) and Livio \& Pringle (1996)
as the same timescale of the precessing period, i.e.,
\begin{eqnarray}
\tau_{\rm prec} & \simeq & 40
\left( {{M_{\rm disk}} \over {1 \times 10^{-6} ~M_\odot}} \right)
\left( {{R_{\rm disk}} \over { 50 ~R_\odot}} \right) \cr
&\times & \left( {{P_{\rm orb}} \over {223 \mbox{~day}}} \right)^{-1}
\left( {{L_{\rm bol}} \over {2 \times 10^{38} \mbox{~erg~s}^{-1}}}
\right)^{-1} \mbox{~day~},
\label{precession_time}
\end{eqnarray}
where we assume that the $\alpha$-parameter 
of the standard accretion disk is $\alpha \sim 0.1$. 
Thus, the growth timescale is short enough to excite warping 
of the accretion disk.
It should be noted here that the heated accretion disk surface,
just after the maximum expansion of the WD photosphere, 
could drive a disk wind because the disk has once been engulfed by 
the WD photosphere.  Such a wind could exert 
an even larger back pressure on the disk than radiation and thus
lead to tilting (e.g., \cite{sch94}; \cite{sch96}).  Its growth 
timescale is much shorter than the timescale given by equation
(\ref{precession_time}).  
\par
\begin{figure}
\epsscale{.9}
\plotone{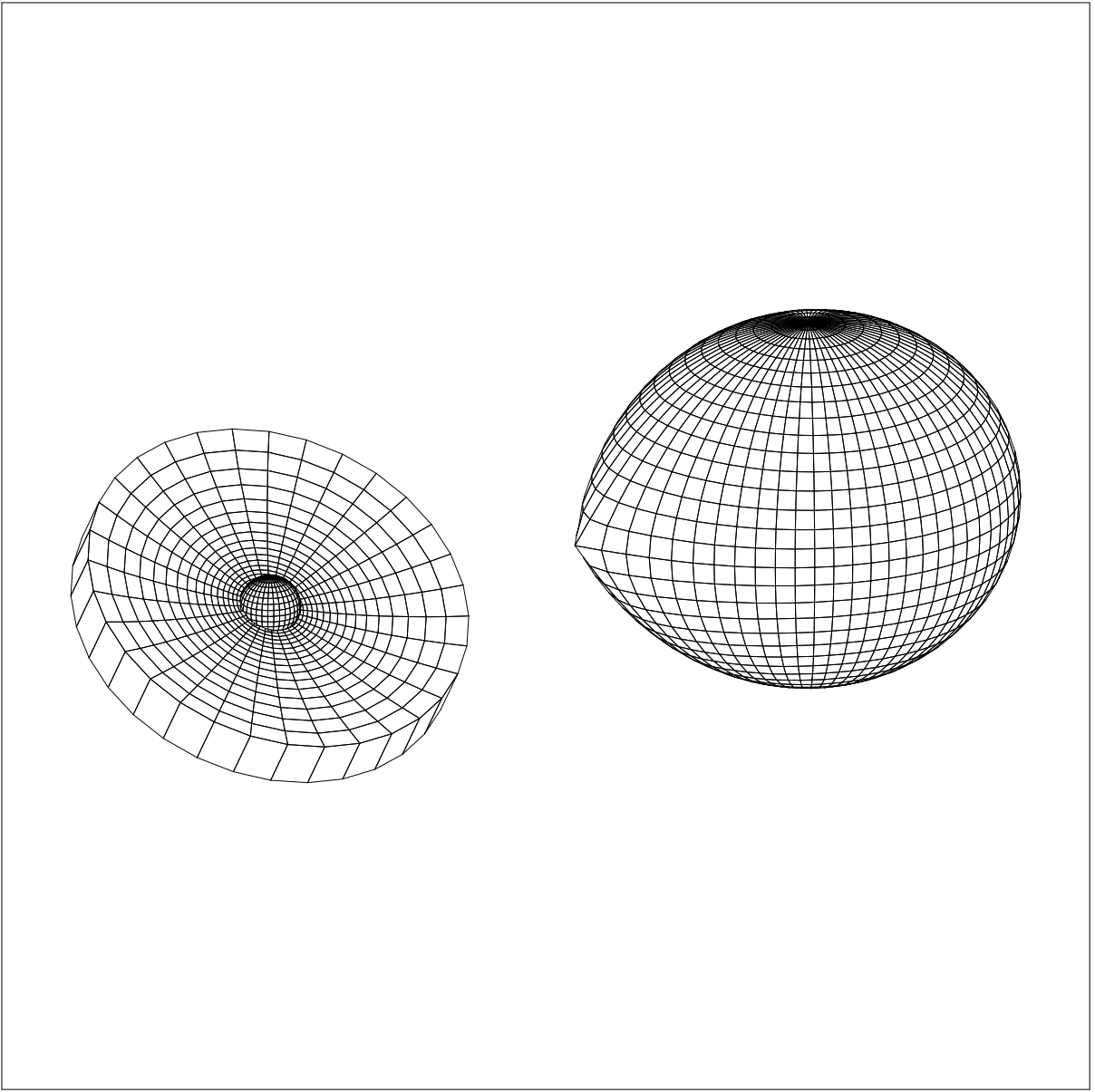}
\caption{
Model configuration near the second peak of the recurrent 
nova T CrB.  The cool component (right in the figure) is a red-giant 
filling up its inner critical Roche lobe.  The hemisphere is 
heated up by the hot component ($1.35 ~M_\odot$ white dwarf, 
left in the figure).  
We assume a tilting accretion disk around the hot component, which is 
precessing about 10\% faster than the orbital rotation.  
The surface of the accretion disk is also heated up by the hot 
component.  The photospheric radius of the hot component 
near the second peak is as small as 
$\sim 0.05 ~R_\odot$, about $\sim 0.001$ times the size 
of the cool component, but 
it is exaggerated in this figure to easily see it. 
\label{vmaxfig}}
\end{figure}
\begin{figure}
\epsscale{.9}
\plotone{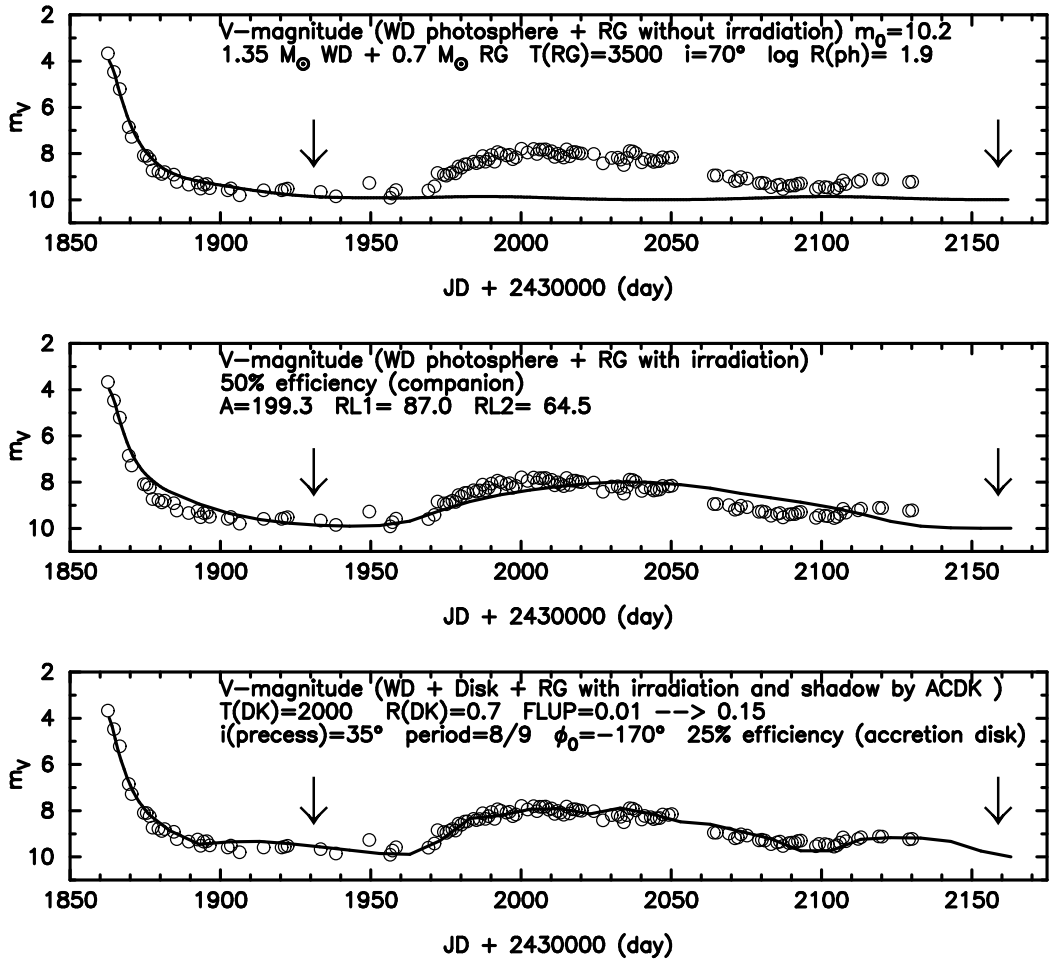}
\caption{
Model light curves are plotted against time (JD 2430000+) 
together with the observational points 
(Pettit 1946a, 1946b, 1946c, 1946d).
Top panel: the visual magnitude of the sum of 
the white dwarf (WD) photosphere and the red giant (RG) 
photosphere without irradiation.  Large arrows 
indicate epochs at the spectroscopic conjunction 
with the M-giant in front. 
Middle panel: the visual magnitude of the WD photosphere and 
the RG photosphere with irradiation.  
Bottom panel: the visual magnitude of the WD, RG with irradiation, 
and the accretion disk heated-up by the hot component.  
\label{vmag1350va1}}
\end{figure}

     The large inclination angle of the accretion disk such as 
$i_{\rm prec} \sim 30-35 \arcdeg$ is required to reproduce 
the observational light curve mainly because we need a 
reflection area as large as the companion, 
which is viewed from ours as shown in Figure \ref{vmaxfig}.
About 10\% faster precessing angular velocity is also required from 
the phase relation between the rising shoulder of the second peak 
near 1970 (JD 2430000+), a small dip near 2100 JD, and then a small 
bump near 2130 JD as shown in Figure \ref{vmag1350va1}.  
These dips are caused by a large shadow on the companion cast 
by the accretion disk.  
This precession velocity is consistent with the fact 
that the radiation-induced precession is prograde, 
while the tidally induced precession is retrograde.  

\acknowledgments
     This research has been supported in part by the Grant-in-Aid for
Scientific Research (08640321, 09640325) 
of the Japanese Ministry of Education, Science, Culture, and Sports.

\begin{deluxetable}{ccc}
\footnotesize
\tablecaption{Periods of wind and nuclear burning phases after 
the peak. 
\label{tbl-1}}
\tablewidth{0pt}
\tablehead{
\colhead{$\dot M_{\rm acc}$} &
\colhead{wind phase} &
\colhead{H-burning} \nl
\colhead{$(10^{-7}~M_\odot \mbox{~yr}^{-1})$} &
\colhead{(day)} &
\colhead{(day)} 
} 
\startdata
0.0 & 75 & 111 \nl
1.0 & 78 & 121 \nl
2.0 & 82 & 135 \nl
4.0 & 93 & $\infty$ \nl
\enddata
\end{deluxetable}

%

\end{document}